# A Pilot Study on Teacher-Facing Real-Time Classroom Game Dashboards

Luke Swanson, David Gagnon, Jennifer Scianna

Field Day Lab, University of Wisconsin - Madison

**Acknowledgements**

This material is based upon work supported by the National Science Foundation under grants DRL-1907384, DRL-1907398, and DRL-1907437. Any opinions, findings and conclusions or recommendations expressed in this material are those of the authors and do not necessarily reflect the views of the National Science Foundation.

**A Pilot Study on Teacher-Facing Real-Time Classroom Game Dashboards**

Teachers in primary and secondary education often make use of educational video games as in-class activities for their students. However, a teacher's only approach to understand their students' gameplay may be to check each player's computer screen individually during game time (Hanghøj & Brund, 2010). This process requires that the game's state can quickly and easily be read from the player screen, and that the teacher has a good understanding of the details of the gam,. A failure in either case can hinder the teacher's ability to provide feedback and guide student progress toward the game's learning objectives.

A real-time presentation of gameplay data on a teacher's own device may offer a solution to this problem. By providing direct insights on each student's gameplay session, we may improve teachers' responsiveness and understanding of gameplay. However, it is not enough to simply dump data onto a screen. Instead, we must present these insights in a manner that can be easily read and understood. This study investigates a few issues related to the design of such a presentation system.

We apply a participatory design method, following the LATUX workflow proposed by Martinez-Maldonado et al. (2015). A similar project is described in (Ruiz et al., 2018). That paper, in particular, set out a similar goal to the project presented here. Namely, the authors used participatory design activities to build a "mental model of the domain expert," which in turn informed the design of a data dashboard. Similarly, Abel & Evans used participatory design to generate ideas for a real-time dashboard for gameplay data (Abel & Evans, 2013).

Several researchers have used other approaches to design teacher-facing dashboard tools for gameplay. Chen et al. investigated how a dashboard could support teachers in their role as a facilitator of a class game session (2020), though they focused on an in-person game, rather than

A Pilot Study on Teacher-Facing Real-Time Classroom Game Dashboards

a computer-based one. Seaton et al. (2019), writing in Data Analytics Approaches in Educational Games and Gamification Systems, discuss a player-facing dashboard implemented for an educational game. Charleer et al. (2018) present an extensive dive into dashboard design for eSports spectators, though this diverges somewhat from teacher use in educational settings. The most extensive work on a dashboard for educational games we could locate was a part of Chaudy & Connolly's recent paper (2018). Their work explores the selection of game data for collection in some depth, and focuses on a broader system for embedding assessments in games..

Clearly, there is some momentum in the research community towards development of game data dashboards. However, to our knowledge, few researchers have followed through to implement and test their designs. For example, Seaton et al's design and evaluation appears to be purely hypothetical, and does not discuss input or feedback from actual users. Similarly, the designs generated by Abel and Evans, Ruiz et al., and Chen et al. have not yet been implemented or tested. Chaudy & Connely's dashboard system was tested; however, it appears to be designed for review of past gameplay sessions, and may not be useful as a real-time aid in facilitating a gameplay session, which is the focus of the present work. Thus, while there is clear research interest in building such data-driven tools for teachers, little work has been done to evaluate the designs in practice.

## Method

**Study Design**

Our study was conducted in two phases. First, we collected data on teachers' expectations for a dashboard for game facilitation during a two-day workshop. Teachers and other stakeholders were involved in several design activities, with the results used directly in generating the design for a prototype dashboard in the second phase. We then implemented the prototype and conducted small-scale pilot testing across several classrooms. The prototype consisted of a set of visualized "models," which were metrics based on player data indicating some aspect of player performance.

Throughout the study, we utilized a participatory design approach. Some techniques in participatory design are nicely summarized in (Sanders et al., 2010). Our design and development process followed the LATUX workflow proposed by Martinez-Maldonado et al. (2015). In particular, the first two stages of LATUX (problem identification and low-fidelity prototyping) were realized through the teacher workshop, while the high-fidelity prototyping and pilot study stages occurred during the second phase of our study. Stage 5 of the LATUX workflow deals with larger, longer-term evaluations of the tool when deployed to the classroom, which we leave for future work.

**Lakeland: A Real-time, City-building Game**

In order to focus the scope of our design work, we chose to build the game dashboard for a specific educational game. The game we selected for this purpose was Lakeland, a real-time strategy, city-building and resource-management game. Lakeland was designed and released by Field Day Lab as an educational game for use in classrooms from grades 7-12. In addition, it was integrated into the Open Game Data system (*Open Game Data*, n.d.), so access to game data logs was readily available.

The choice of game has a significant impact on what gameplay data are available and useful. Thus, we will briefly summarize gameplay and key mechanics. In Lakeland, players are tasked with the development of a farming community. A series of tutorials introduce the main game controls. A player can construct houses, corn fields, and dairy farms on land tiles in their game map. Money is earned by selling produce, although players must reserve some of the food they produce to keep their farmers alive. A "nutrition" overlay allows players to see the nutrient concentrations across their map tiles; cornfields require (and rapidly deplete) nutrients as they grow, requiring the player apply fertilizer to ensure regular harvests. On the other hand, a high nutrient concentration in lake tiles will lead to a toxic algae bloom, a problem exacerbated as rain washes excess fertilizer towards lakes.

**Figure 1**

*A Sample Image of the Lakeland Game*

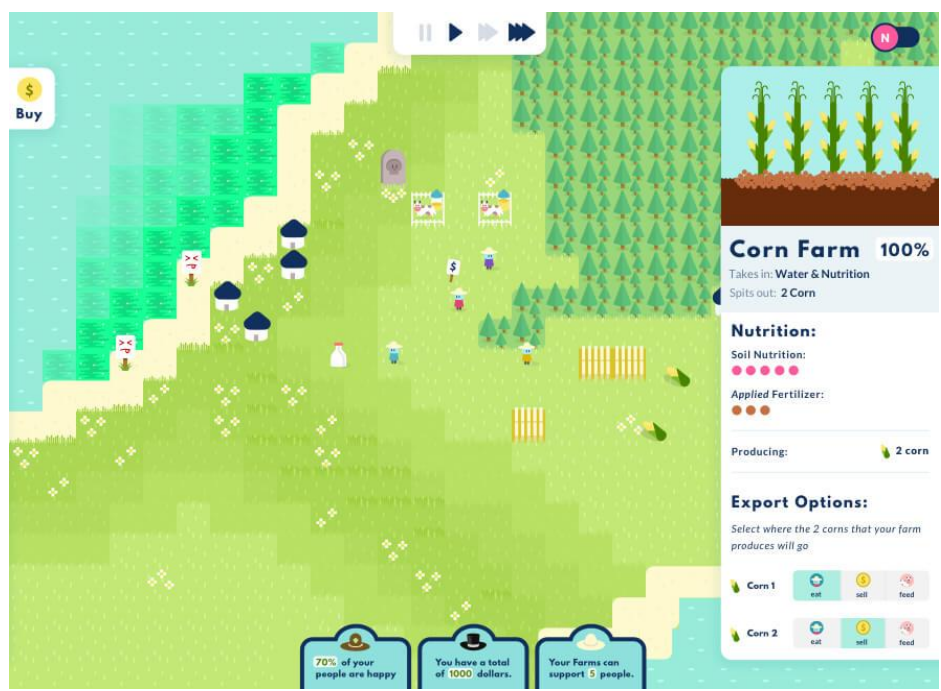

Thus, the core strategy of the game centers around growing a community while maintaining a balance of money, food, and nutrients. A gameplay approach focused purely on maximizing profits and farm growth will eventually fail due to the environmental toll on the community's lakes. One particular quirk of the game's "fertilizer runoff" mechanic causes fertilizer to move on a diagonal toward the nearest lake. Players can improve their strategy by placing crop fields diagonally, ensuring that fertilizer will run across other fields on that diagonal en route to the lake. As a player progresses through the game, they can earn achievements based on number of residents, number of fields, money earned, and number of algae blooms.

**Participatory Design Workshop**

We began the first phase of this study with a workshop involving teachers and other project stakeholders. This event featured discussions of game data and the use of games in the classroom generally, as well as design exercises meant to elicit user requirements for our dashboard design. We recruited twelve teachers and "teaching mentors" to participate in the workshop and related design activities. We consider teachers to be experts on their classrooms and on facilitation of classroom activities; thus this phase of the study was intended to capture insight into their expertise.

The "design" portion of the workshop consisted of three cumulative activities, which resulted in mockups of dashboard designs from the teachers themselves. In the first activity, participants were asked to separate into pairs, and received a stack of "prompt" cards to fill out during the design time. There were three types of card, each with a particular prompt for one of three types of game data model, namely detectors of a specific behavior, monitors of general gameplay, and predictors of future decisions. All card prompts followed a basic format of: "When I facilitate a game play activity, I would like to __________ so that I can respond by __________." The card types varied slightly in the "I would like to" clause (see Figure 2). For the first design activity, the teachers were instructed to work with their partners to fill out the prompts based on the features of gameplay they wanted to see in a dashboard tool.

In the second design activity, teachers were asked to fill in the right sides of the cards they completed during the first activity. The right sides included spaces for the participants to draw (or describe) their idea of a visual representation for the model described on the left. These

A Pilot Study on Teacher-Facing Real-Time Classroom Game Dashboards

visualizations were intended to form a basis for the third activity. Finally, the teachers were given large paper sheets and asked to use their model visualization ideas from the first two activities to design a full dashboard mockup. For this activity, teachers were allowed to work individually or in pairs.

**Figure 2**

*Completed Prompt Cards from First and Second Design Activities*

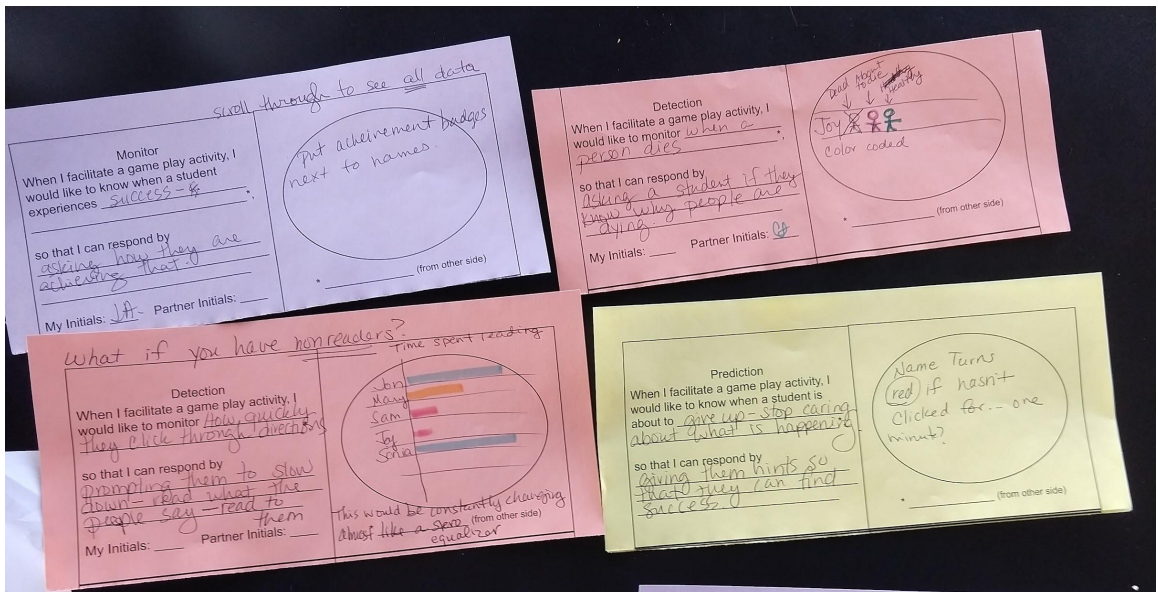

All prompt cards and dashboard mockups from these design activities were collected as design artifacts, and analyzed to generate insights for development of our dashboard prototype. Our primary concern for this analysis was the content of the prompt cards, which offered the most direct insight into the concepts educators wanted the dashboard to communicate. We tabulated the prompt cards and recorded the responses to each "blank" in the prompts, along with a simple description of the visualizations, if any. We coded responses to the "I would like to know/monitor" blanks, with codes reflecting the kind of metric or model the writer had asked for. From this, we generated counts for each metric category, which indicated what models generated the most interest from teachers.

A Pilot Study on Teacher-Facing Real-Time Classroom Game Dashboards

**Dashboard Prototyping & Pilot Study**

The second phase of our study consisted of development of a working dashboard prototype, and a pilot study using the prototype in authentic class contexts. We briefly summarize the dashboard system, before describing the pilot study itself. Pilot data was collected through post-session surveys and interviews.

*System Design*

The dashboard prototype system has client-side and server-side components. When a student plays Lakeland, their game session sends event log data to be stored in a database. The server-side portion of the dashboard system retrieves game logs from the database, and extracts our predefined gameplay features from the data. The client-side code runs as a webpage on a teacher's device, sending periodic requests for updated feature values to the server, and transforming the resulting values into visualizations. This architecture is summarized in Figure 3.

**Figure 3**

*Architecture of Prototype Dashboard Tool*

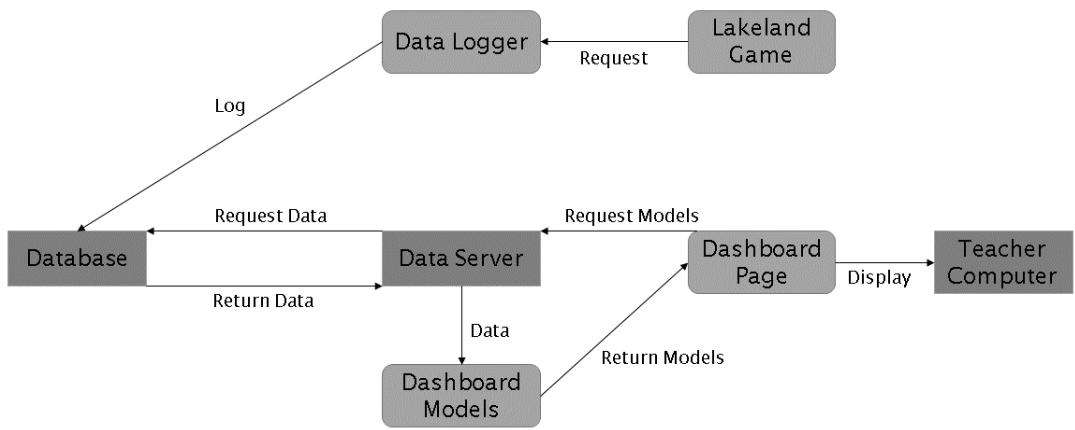

*Note.* Darkened rectangles represent separate physical devices, while rounded rectangles represent software components.

A Pilot Study on Teacher-Facing Real-Time Classroom Game Dashboards

In order to show a teacher only the players in their class, a "portal" page was used for accessing the Lakeland game. When a teacher opens the dashboard page, they are asked to enter a class code, which generates a portal link the teacher shares with their students. Students enter a user name in the portal page, and are redirected to the Lakeland game itself. In this way, the dashboard system can request data only from sessions associated with the teacher's classroom code, and each player in the class can be properly labeled in the dashboard. This allows teachers to identify individuals in their dashboard, while maintaining student privacy from people other than their teacher.

*Classroom Tests*

After developing our initial prototype, we conducted a short pilot study with teachers. The goal of this test was to assess whether the prototype, built through a participatory design process, was able to play a useful role in teachers' classroom game sessions. Prior to classroom testing, each teacher was given instructions for how to use the system, and scheduled a day for playtesting. Following their tests, each teacher was asked to fill out a short post-session feedback survey, and to schedule an exit interview, both within 24 hours of the first classroom session. This was intended to help standardize the feedback and interviews, particularly in cases where teachers tested with multiple classes.

The survey and interview questions were developed to help us address the following three research questions: First, did teachers' inputs clearly influence the prototype design? That is, was the tool co-designed? Second, is the prototype itself usable in the context of classroom gameplay? That is, is the tool feasible? Third, does the prototype provide useful information, and improve the teachers' experience? That is, is the tool effective?

The survey included general information about the classroom test, such as the teacher's name, test context (face-to-face vs. online), and number of students. Other sections of the survey asked for feedback on specific parts of the user experience, namely co-design, feasibility, and efficacy. For each of these three categories, the poll contained a three to five 5-point Likert-scaled questions, and three free response questions.

Finally, each interview was between 30 minutes and one hour long. As with the survey, interview prompts attempted to capture teacher insights into the results of co-design, as well as the prototype's feasibility and efficacy, with some variation due to individual teachers' circumstances. For example, a teacher who led only online sessions would not be able to answer prompts about how the process with the dashboard differed from their "normal" approach in classes, because an online session already fundamentally differs from "normal" face-to-face sessions.

## Results

**Insights From Design Workshop**

We now discuss our analysis of the "prompt cards" produced during the design activities of our teacher workshop. We collected and analyzed 71 prompt cards in total, coding teacher responses to the prompts, to generate categories of data models. In total, we identified 43 such categories. However, most categories appeared only on one card. To narrow the scope of choices, we considered only the 15 categories represented on multiple cards for inclusion in the initial prototype. These are listed in Table 1.

A Pilot Study on Teacher-Facing Real-Time Classroom Game Dashboards

**Table 1**

*Model Categories Appearing on Multiple Prompt Cards*

| Category | Count | Category | Count | Category | Count |
|---|---|---|---|---|---|
| Frustration | 6 | Success | 4 | House | 2 |
| Strategy | 5 | Indecision | 3 | Understanding | 2 |
| Progress | 4 | Give Up | 3 | Buy | 2 |
| Time | 4 | Farm | 3 | Stuck | 2 |
| Deaths | 4 | Build | 3 | Achievement | 2 |

Several categories, including game events like buy, build, and deaths can be shown directly in a dashboard. On the other hand, some of the most popular categories deal with abstract concepts like frustration, strategy, and success. These may not have direct representations in the game data. From these most-requested categories, we selected seven to develop into models for the dashboard pilot study. These were Progress, Strategy, Death, Frustration, Give Up, Indecision, and Game State. Based on the seven categories highlighted here, we developed 16 data models, which were implemented in the dashboard prototype. These are listed in Table 3.

We should note that frustration, giving up, and indecision are three similar but distinct categories. For this analysis, an indecisive player has not given up on the game, but does not know how to proceed and may exhibit behaviors similar to a player that has given up. A frustrated player may be on a path to giving up, but is still engaged with the game. The goal in differentiating these three is to analyze data and game mechanics from slightly differing perspectives, in order to find models that can clue teachers in to different patterns of behavior consistent with each category. Finally, game state as a category is intended to capture "farm,"

A Pilot Study on Teacher-Facing Real-Time Classroom Game Dashboards

"build," "house," and similar categories that deal with the structure of players' farm communities.

**Table 2**

*List of Dashboard Prototype Models*

| Model | Category | Visualization |
| --- | --- | --- |
| Tutorials Completed | Progress | Number |
| Tutorial Achievement Rank | Progress | Percentile |
| Money Achievement Rank | Progress | Percentile |
| Bloom Achievement Rank | Progress | Percentile |
| Farm Achievement Rank | Progress | Percentile |
| Population Achievement Rank | Progress | Percentile |
| Playing Time | Game State | Minutes:Seconds |
| Population | Game State | Icon + Number |
| Map Summary | Game State | Bitmap |
| Town Composition | Strategy | Icons + Numbers |
| Diagonal Field Strategy Detector | Strategy | Binary Indicator |
| Time Since Active | Indecision | Binary Indicator |
| Time Since Last Building | Indecision | Binary Indicator |
| Time Since Last Sale | Frustration | Binary Indicator |
| Time Since Tile Exploration | Give Up | Binary Indicator |
| Farmer Deaths | Death | Icon + Number |

**Classroom Pilot Study**

We recruited teachers from the original set of workshop participants to join in the pilot study. Four teachers responded and followed up to schedule a classroom test. Three of the four teachers tested with multiple classes, with one teacher reporting that they had 12 different class sessions during which they intended to test the dashboard prototype. Classrooms tested ranged from 6th to 12th grade, with no more than 20 students per session. Due to coronavirus-related school shutdowns, only one teacher was able to conduct their test in a face-to-face classroom; the other instructors conducted online classroom play sessions.

Given the small sample size, the results of the post-session survey can not give strong empirical conclusions, but allowed us to prompt for feedback in a structured, standardized way. The results provide a starting point for considering the prototype's success in each case. Averages from the teacher responses are in tables 3, 4, and 5. Answers to the free-response prompts and interviews offer deeper insights and context for the scores.

### *Evidence of Co-design*

When teachers participated in the pilot study, they were encouraged to reflect on aspects of the prototype they felt reflected their input. In the post-gameplay survey, there was a strong response in terms of evidence of a teacher's own input in the prototype design, and in helping teachers to understand what changes could improve the tool. In the free responses, one teacher mentioned that they remembered discussing the diagonal strategy detector during the original workshop. Another felt that the overall "look of the dashboard is what many shared" in the workshop designs. A third noted that "the dashboard helps educators give students feedback and guide their gameplay." This last seems to indicate that the core design of the dashboard prototype aligns well with teachers' goals for the tool. Regarding features the teachers perceived to be missing from this first iteration of the dashboard, some referenced the lack of overview data, though class-level overviews were excluded from design due to project scope

**Table 3**

*Summary of Survey Results for Evidence of Co-design*

| Co-designed | |
|---|---|
| Prompt | Average Score |
| My feedback and participation is evident in the dashboard prototype | 4.5 |
| Testing this tool helped me understand how the next version should be changed | 4.25 |
| This tool helped me understand the Lakeland game itself | 3.75 |

*Feasibility*

The feasibility scores from the survey are generally high. Teachers seemed to appreciate the presence of the dashboard when checking on individual students. From one interview: "I have a lot of students who might disengage or just click randomly. And I feel... with a dashboard, it's useful to know [that] they're actually making progress, or what [they are] making progress on." One said they cycled through the student dashboards, often checking the map preview first to understand how a student was doing in the game. In their words "[in] the first five minutes of the game, it doesn't change that much. But every time I opened up a new student, I would always [look] there first and [ask] 'is it covered with farms now'?" This teacher also noted that the dashboard may be a better fit in the classroom if it contained a class overview, or better ways to see which students might need help without cycling through the players. Another described a sort of inverse process. They spent time checking student screens, and then turned to the dashboard to help understand what was going on in that particular game. A third was grateful to be able to "sit back and watch the dashboard... instead of me having to watch their screen." The fourth teacher stated that they spent some time projecting the dashboard for their class to see, and noted their students enjoyed seeing their statistics on screen.

However, it appears the prototype did not always work as teachers expected. One in particular noted an issue with their class code led to the prototype failing during their first attempt. Another two teachers noted that a few sessions appeared in their dashboard which were not associated with a student in their class. However, the issues noted here seem to deal more with specific bugs in the prototype, not with the feasibility of a dashboard for use in classrooms. One teacher felt that there may have been too much data on the screen at once to work at a quick glance, which is a design consideration worth further evaluation.

**Table 4**

*Summary of Survey Results for Feasibility*

| Feasibility | |
| --- | --- |
| Prompt | Average Score |
| The tool functioned as expected | 3.75 |
| Setting up a class code and getting students connected was easy | 5 |
| The dashboard itself was easy to use | 4 |
| If I were to play Lakeland again, I would use this tool | 5 |
| I will recommend my colleagues use this tool when they use Lakeland | 5 |

***Effectiveness***

Regarding the prototype's effectiveness, the survey numbers again indicate issues with the prototype in certain cases, but are positive overall. In the free response questions for "effectiveness," teachers were specifically prompted for instances where the dashboard appeared to be inaccurate. The issues reported here were similar to those from the previous section. One teacher indicated that their students did not appear in the dashboard on the first run, and another stated some players disappeared mid-game, and failed to reappear.

On the other hand, there was a positive response to the general design. "I like the fact that it had icons and colors to it... it was organized in a nice layout." Some teachers were able to describe instances where the prototype specifically helped them to identify struggling students; one used the "inactive" alert to help them identify online students with technical/connection issues, while another had "one student who [was] inactive, and I noticed that he had zero out of six [tutorials] done," which helped them to assist that student. Another appreciated the indicator "light" for the diagonal strategy detection. "I had a couple students today that actually, like that light bulb went green... they figured it out, and I was excited."

**Table 5**

*Summary of Survey Results for Effectiveness*

| Effectiveness | |
| --- | --- |
| Prompt | Average Score |
| The tool provided useful insights to my students' play | 4 |
| I trust the data presented | 4.25 |
| This tool helped me facilitate the game session | 3.5 |
| The dashboard interface communicated quickly and clearly | 4.25 |

However, further work is needed for some of the models and their visualizations, as teachers indicated trouble understanding certain models. One was unclear on the meaning of the percentile indicators. Another liked the percentiles, but "I would [prefer to] group those together, or make them more visual."

## Discussion

In this study, we followed a participatory design process. Prompt cards, filled out by individual teachers during a design workshop, allowed us to capture expert insight into the needs of teachers facilitating classroom gameplay. As demonstrated through the discussion of workshop results and dashboard model selections, these cards provided a clear path to development of dashboard models.

Teacher responses to the post-session survey and interviews were generally positive. The respondents were able to identify evidence their original input contributed to the prototype we developed. They also praised several of the individual models in the dashboard for helping them to understand what was happening in a given game session.

On the other hand, there were some issues with the prototype. The scope of design must be expanded to include class-level overviews and summary data in the player list. While teachers generally found the classroom portal system easy to use, they discussed some bugs that limited effectiveness and ease of use. However, despite these issues, teachers clearly indicated that this kind of dashboard tool has a role in supporting their facilitation of gameplay. It is clear this project was successful in producing a tool that already benefits teachers. Participatory design, shaped by judicious application of established design principles, offers a reasonable method for co-developing classroom gameplay tools with and for teachers.

This study demonstrates in detail the process used for a successful pilot project. The tool we developed aided teachers in understanding both a given game and the students who played it. Future work on these tools may enable teachers to better implement gameplay in their classrooms.

A Pilot Study on Teacher-Facing Real-Time Classroom Game Dashboards

A Pilot Study on Teacher-Facing Real-Time Classroom Game Dashboards